# Semitransparent anisotropic and spin Hall magnetoresistance sensor enabled by spin-orbit toque biasing

Yumeng Yang[1], Yanjun Xu[1,2], Hang Xie[1], Baoxi Xu[2], and Yihong Wu[1,a)]

[1]*Department of Electrical and Computer Engineering, National University of Singapore, 4 Engineering Drive 3, Singapore 117583, Singapore*

[2]*Data Storage Institute, A*STAR (Agency for Science, Technology and Research), 2 Fusionopolis Way, 08-01 Innovis, Singapore 138634, Singapore*

We demonstrate an ultrathin and semitransparent anisotropic and spin Hall magnetoresistance sensor based on NiFe/Pt heterostructure. The use of spin-orbit torque effective field for transverse biasing allows to reduce the total thickness of the sensors down to 3 - 4 nm and thereby leading to the semitransparency. Despite the extremely simple design, the spin-orbit torque effective field biased NiFe/Pt sensor exhibits level of linearity and sensitivity comparable to those of sensors using more complex linearization schemes. In a proof-of-concept design using a full Wheatstone bridge comprising of four sensing elements, we obtained a sensitivity up to 202.9 m$\Omega$ $Oe^{-1}$, linearity error below 5%, and a detection limit down to 20 nT. The transmittance of the sensor is over 50% in the visible range.

a) Author to whom correspondence should be addressed: elewuyh@nus.edu.sg

Transparent sensors offer possibilities for emerging applications in internet-of-things (IOT) and smart living. Although a variety of transparent or semitransparent devices have been demonstrated using semiconductors,[1,2] polymers,[3,4] two-dimensional materials,[5-7] *etc.*, it remains a great challenge to achieve the same in magnetic devices. This is because most of the practical magnetic materials are metals whose transmissivity in the visible range of electromagnetic spectrum diminishes quickly as the thicknesses increases. For instance, the transmittance of Fe, Co and Ni is only about 20% at a thickness of 10 nm, and it decreases to about 5 - 6% at 20 nm. As for most magnetic sensors, in addition to the ferromagnetic active layer, one also needs additional layers for magnetic biasing which is essential for sensor linearization, and its total thickness can easily exceed 20 nm.[8,9] This makes it difficult, if not impossible, to realize all-metal-based transparent magnetic sensors using the conventional bias schemes.

Here we report a semitransparent anisotropic and spin Hall magnetoresistance (MR) sensor with a transmittance exceeding 50% in the visible range. The key to achieving semitransparency is the use of spin-orbit torque (SOT) effective field for transvers bias which significantly reduces the total thickness of the sensor, down to 3 - 4 nm. The SOT has been reported in a variety of ferromagnet (FM) / heavy metal (HM) heterostructures since its first observation in Pt/Co/$AlO_x$.[10] Although the exact mechanism is still being debated, it is generally accepted that two types of torques are present in the FM/HM heterostructures, one is called field-like (FL) and the other is (anti)damping-like (DL). Phenomenology, the two types of torques can be modelled by $\vec{T}_{DL} = \tau_{DL}\vec{m} \times [\vec{m} \times (\vec{j} \times \vec{z})]$ and $\vec{T}_{FL} = \tau_{FL}\vec{m} \times (\vec{j} \times \vec{z})$, respectively, where $\vec{m}$ is the magnetization direction, $\vec{j}$ is the in-plane current density, $\vec{z}$ is the interface normal, and $\tau_{FL}$ and $\tau_{DL}$ are the magnitude of the FL and DL torques, respectively[11-13]. It should be pointed out that the sign of $\vec{z}$ depends on the sign of the HM spin Hall angle and on whether the top or the bottom HM interface is in contact with the FM; and therefore the directions of the effective fields and torques will be opposite if the stacking order is reversed. If $\vec{m}$ does not change significantly, the two toques can be expressed in the form of $\vec{M} \times \vec{H}_{eff}$, where $\vec{H}_{eff}$ is an effective field. Following this notion, the FL effective field ($H_{FL}$) is in the direction of $\vec{j} \times \vec{z}$, whereas the DL effective field ($H_{DL}$) is in the direction of $\vec{m} \times (\vec{j} \times \vec{z})$. Since $H_{FL}$ is independent of the magnetization and transverse to the charge current, it naturally functions as a transverse bias for a MR sensor with in-plane magnetic anisotropy (IMA). On the other hand, for films with IMA, the effect of $H_{DL}$ to drive the magnetization out-of-plane is negligible due to the large out-of-plane demagnetizing field ($H_{Dz}$), *e.g.*, for a NiFe(1.8)/Pt(2) ellipsoid with a lateral dimension of 800 μm × 200 μm, the experimentally determined $H_{Dz}/H_{DL}$ ratio is around 1000 at a current density of $j_{Pt}$ = 7.34 $\times 10^5$ A $cm^{-2}$. In conventional designs of anisotropic magnetoresistance (AMR) sensors, in addition to the active sensing layer, there is always a need for additional layer to provide the transverse bias field, such as a soft-adjacent layer (SAL) in SAL-biasing and patterned conductor strips in barber pole biasing,[9] which significantly increases the complexity in sensor design and manufacturing. In addition, the total thickness of the sensor can easily exceed 20 nm. The use of SOT effective field for biasing does not only simplify the sensor structure but also renders it possible to make semitransparent sensors. Although the concept of SOT biasing applies to different FM/HM combinations, here we focus on sensors based on NiFe/Pt bilayers. By optimizing the individual layer thicknesses, we obtained a full Wheatstone bridge MR sensor with a linearity error below 5%, sensitivity up to 202.9 mΩ $Oe^{-1}$ and transmittance over 50%.

The transmittance of FM/HM bilayers can be readily calculated using the transfer matrix method.[14] The inset of Fig. 1(a) shows a typical sensor structure consisting of a HM layer, a NiFe layer and supporting substrate. The thicknesses and refractive indices of the individual layers are $d_m$ ($m$ = 1 for FM and 2 for NiFe), $d_S$ and $n_m$, $n_S$, respectively. Here, the reflective indices are in general complex numbers. We also assume that $d_s$ approaches infinity. Assuming that light travels in the *zx* plane with either *s*-polarization ($\vec{E} \parallel \hat{y}$) or *p*-polarization ($\vec{H} \parallel \hat{y}$), the amplitude of the electrical field of a plane wave that satisfies the Maxwell equation can be written as $E = [A(x) + B(x)]e^{i(\omega t - k_z z)}$, where $k_z$ is the $z$ component of the wave vector, $\omega$ is the angular frequency, $t$ is time, and $A(x)$ and $B(x)$ are amplitude of the right-travelling and left-travelling waves, respectively. The amplitude of the electrical field inside the air and those after passing through the $m^{th}$ layer and substrate interface are related by the following equation:

$$\begin{pmatrix} A_0 \\ B_0 \end{pmatrix} = D_0^{-1}[\prod_{m=1}^{N} D_m P_m D_m^{-1}] D_s \begin{pmatrix} A_s \\ B_s \end{pmatrix} \quad (1)$$

where

$$D_m = \begin{pmatrix} 1 & 1 \\ n_m \cos\theta_m & -n_m \cos\theta_m \end{pmatrix} \quad \text{for } s\text{-polarization}$$

$$D_m = \begin{pmatrix} \cos\theta_m & \cos\theta_m \\ n_m & -n_m \end{pmatrix} \quad \text{for } p\text{-polarization} \quad (2)$$

and $P_m = \begin{pmatrix} e^{i\omega_m} & 0 \\ 0 & e^{-i\omega_m} \end{pmatrix}$ is the propagation matrix, $\omega_m = \frac{2\pi n_m d_m cos\theta_m}{\lambda}$ is the change in phase after the light passing through the $m^{th}$ layer. Here, $\lambda$ is the wavelength, and $\theta_m$ is angle of incidence in the $m^{th}$ layer. If we let $D_0^{-1}[\prod_{m=1}^{N} D_m P_m D_m^{-1}] D_s = \begin{pmatrix} M_{11} & M_{12} \\ M_{21} & M_{22} \end{pmatrix}$, then the transmittance is given by $T = \frac{n_s cos\theta_s}{n_0 cos\theta_0}\left|\frac{1}{M_{11}}\right|^2$. For unpolarized light, one can take an average of the contributions from both the $s$-polarization and $p$-polarization light. Fig. 1(a) shows the simulated transmittance ($\theta_m$ = 0) in the visible range for NiFe(1.5)/HM(2) bilayers on quartz substrate with different HMs, *i.e.*, Pt, Ta and W (number in the parenthesis indicates thickness in nanometer). It is clearly seen that all the bilayers exhibit a transmittance over 50%. The different trend of the curves for different HMs is due to the different dispersion of refractive indices. The transmittance can be further enhanced by adding an oxide capping layer that functions as an anti-reflection coating. Fig. 1(b) shows the simulated oxide thickness dependence of transmittance for Pt(2)/NiFe(1.5)/oxide($d_{oxide}$) trilayers [see inset of Fig. 1(b)] with different oxides $Ta_2O_5$, $SiO_2$, MgO and $Al_2O_3$ at $\lambda$ = 500 nm. As a reference, the transmittance of Pt(2)/NiFe(1.5) bilayer is also shown in dashed line. With the oxide anti-reflection coating, it is possible to achieve a transmittance up to 70% capped by $SiO_2$, MgO and $Al_2O_3$ layers. It should be noted that, in addition to enhancement of transmittance, the oxide capping layer may also help to strengthen the SOT effect as reported in literatures.[15,16] As a proof-of-concept experiment, in this work we only focus on the experimental results obtained in the NiFe/Pt bilayers.

The NiFe/Pt bilayers were deposited on quartz substrates with the NiFe layer deposited first by e-beam evaporation and followed by the deposition of Pt using DC magnetron sputtering. Both layers were deposited in a multi-chamber system at a base pressure below $3\times10^{-8}$ Torr without breaking the vacuum. An in-plane field of ~500 Oe was applied during the deposition to induce a uniaxial anisotropy for the magnetic film. Before patterning into sensor elements, thickness optimization was carried out on coupon films by characterizing both the optical transmittance and magnetic properties. Fig. 1(c) shows the measured transmittance for NiFe(1.5)/Pt($d_{Pt}$) films with $d_{Pt}$ = 1.5 nm, 2 nm and 2.5 nm, respectively. As a reference, the transmittance of bare quartz substrate is also shown in the figure. The measured transmission spectra are in good agreement with the simulated results shown in Fig. 1(a); and as expected, over 50% transmittance is obtained for the $d_{Pt}$ = 2 nm sample in the visible range. As shown in the inset of Fig. 1(c), the semitransparency of the NiFe(1.8)/Pt(2) bilayer is clearly demonstrated in the photograph of NUS logo covered by the coupon film. The magnetic properties of the films were characterized by measuring the *M-H* loops using a vibrating sample magnetometer with field applied in-plane in the induced anisotropy axis direction. The results are shown in Fig. 1(d) for NiFe($d_{NiFe}$)/Pt(2) with $d_{NiFe}$ = 1.7 nm, 1.8 nm, 1.9 nm, and 2 nm, respectively. Both the $d_{NiFe}$ = 1.9 nm and 2 nm samples exhibit typical soft FM behavior with in-plane anisotropy and a coercivity of around 1 Oe, whereas the $d_{NiFe}$ = 1.7 nm sample shows a superparamagnetic behavior at room temperature. The behavior of the $d_{NiFe}$ = 1.8 nm sample falls between those of $d_{NiFe}$ = 1.7 nm and 1.9 nm: it shows a clear magnetization switching but negligibly small coercivity. It is possible that a small portion of the sample becomes superparamagnetic while the remaining part is FM. In view of these results, in order to harness the high transmittance and large SOT effect at small thickness yet not to compromise the FM behavior, we chose to fabricate SOT-biased sensors with a structure of NiFe(1.8)/Pt(2).

Figure 2(a) shows the schematic of a full Wheatstone bridge MR sensor consisting of four ellipsoidal NiFe/Pt bilayer sensing elements for differential sensing of an external magnetic field ($H_y$). The ellipsoidal sensing elements were patterned using combined techniques of photolithography and lift-off process with different dimensions. The ratio of long axis length ($a$) over short axis length ($b$) is fixed at $a/b$ = 4, with $a$ = 800 μm, 400 μm and 200 μm, respectively. The electrical contacts (not shown in the schematic drawing) occupies one-third each from the two-ends of the sensor element and, therefore, only 1/3 of the sensor element at the center portion is active for sensing. As illustrated in Fig. 2(b), when the electrical current ($I$/2 each) passes through the two sensor elements [corresponding to 1 and 2 in Fig.2(a)], the effective field $H_{FL}$ is generated in opposite directions and thereby pushing the magnetization, one in upward and the other in downward direction, by an angle $\varphi$ from the longitudinal direction. The same also applies to the other pair of sensor elements, 3 and 4, as shown in Fig. 2(a). Although the Oersted field ($H_{Oe}$) from the current is also in the same direction, its magnitude is generally much smaller compared to $H_{FL}$. A linear response with maximum output ($V$) will be obtained from the other two terminals of the bridge if $\varphi$ can be set at 45$^{o}$ at zero external field. In order to calculate $\varphi$, we must first find $H_{FL}$ as a function of current density.

As we mentioned in the introduction, both DL and FL torques are present in FM/HM heterostructures. It is generally believed that both spin Hall effect (SHE)[17-20] and Rashba-Edelstein (RE)[10,21-23] interaction contribute to the generation of SOT, but their respective roles in the two types of torques are still debatable. Although the RE interaction is expected to play a more dominant role in generating FL torque, recently there is growing evidence to suggest that the FL torque is also attributed to SHE.[20,24-2627,28] Recently, Nan *et al.*[24] have introduced an effective spin Hall angle $\theta_{FL}$ for NiFe/Pt bilayer and express the FL effective field to current density ratio as $\frac{H_{FL}}{j_{Pt}} = \frac{\hbar}{2e}\frac{\theta_{FL}}{\mu_0 M_s d_{NiFe}}$. Based on the reported $\theta_{FL}$value of 0.024 and $M_s$ value of 300 – 500 emu cm$^{-3}$ for ultrathin NiFe, the $H_{FL}/j_{Pt}$ ratio is estimated to be in the range of 0.69 – 1.12×10$^{-6}$ Oe (A$^{-1}$ cm$^2$). In order to quantify the $H_{FL}/j_{Pt}$ ratio experimentally for samples grown on quartz substrate, we measured the $H_{FL}$ for NiFe(1.8)/Pt(2) as a function of current density using the 2$^{nd}$ order planar Hall effect (PHE) method.[29,30] Details of measurement procedure can be found in our previous work.[31] As summarized in Fig. 2(c), the $H_{FL}$ value in the bilayer scales linearly with the current density in Pt layer, and a $H_{FL}/j_{Pt}$ ratio of 0.71×10$^{-6}$ Oe (A$^{-1}$ cm$^2$) is obtained, which is very close to the value obtained from the effective spin Hall angle reported by Nan *et al.*[24] In deriving the $H_{FL}/j_{Pt}$ ratio, $j_{Pt}$ was calculated using the parallel resistor model and experimentally obtained resistivity values: $\rho_{Pt}$ = 31.66 μΩ·cm and $\rho_{NiFe}$ = 78.77 μΩ·cm. Note that the contribution from $H_{Oe}$ has already been subtracted out from the $H_{FL}$ values shown in Fig. 2(c) by using the relation $H_{Oe} = d_{Pt} j_{Pt}/2$, where $d_{Pt}$ is the thickness of Pt. With this $H_{FL}/j_{Pt}$ ratio, we can then proceed to calculate the current density that is required to set $\varphi$ at 45$^o$ at zero external field.

Based on the macro-spin model, the free energy density $\varepsilon_{Tot}$ of the sensor element is given by[32]

$$\varepsilon_{Tot} = \frac{\mu_0}{2}\cdot M_s^2 N_x + \frac{\mu_0}{2}\cdot M_s\left[M_s\left(N_y - N_x\right) + H_k\right]\cdot\sin^2\varphi - \mu_0 M_s H_{bias}\sin\varphi \quad (3)$$

where $H_{bias} = H_{FL} + H_{Oe}$, $N_x$ and $N_y$ are the demagnetizing factors in *x*- and *y*-direction, respectively, and $H_k$ is the anisotropy field. Here, $N_x$ and $N_y$ are calculated by approximately treating the ellipsoid in a rectangular shape with a dimension of $a\times b\times t_{NiFe}$.[33] By taking $\mu_0 M_s$ = 0.42 T and $H_k$ = 0.60 Oe (extracted experimentally), $H_{bias}$ that is required to bias the sensor at different angle $\varphi$ can be obtained by minimizing the energy density, which can then be converted to the current density in Pt. The results are shown in Fig. 2(d) for sensors of different dimensions: $a$ = 800 μm, 400 μm and 200 μm, respectively. The dashed lines indicate $\varphi = \pm 45^{o}$. It can be seen that in all cases, the magnetization can be biased to $\pm 45^{o}$ with a current density in the range of 0.8 - $1.2\times10^{6}$ A $cm^{-2}$, which is reasonable for normal magnetic sensor operation. Once a proper biasing in obtained, the output voltage of a single sensing element is given by $\Delta V = \frac{\Delta\rho_{NiFe}}{\rho_{NiFe}}\frac{a\rho_{Pt}j_{Pt}}{3}\frac{\sqrt{2}H_y}{H_k+H_d}$. Here, $\Delta\rho_{NiFe}$ is the change in resistivity of the NiFe layer caused by $H_y$. In a single layer of NiFe, $\Delta\rho_{NiFe}$ mainly comes from the AMR effect. However, in the case of NiFe/Pt bilayer, in addition to AMR, spin Hall magnetoresistance (SMR) also contributes to the resistivity change induced by the external field. In fact, for NiFe(1.8)/Pt(2) bilayer, we found experimentally that the contribution of SMR is two times as large as that of AMR. Despite the different origins, both the AMR and SMR follows the same angle-dependence: $\Delta\rho = (\Delta\rho_{AMR} + \Delta\rho_{SMR})cos^2\varphi$ ,[34,35] where $\Delta\rho_{AMR}$ ( $\Delta\rho_{SMR}$ ) represents the size of resistivity change induced by AMR (SMR). This allows one to lump both effects together and ensures that SOT-biasing works for both AMR and SMR sensors.

We now turn to the typical performance of SOT-biased sensors in detecting both DC and low-frequency AC field generated by a pair of Helmholtz coils. In order to reduce the influence of earth field, both the sensor and Helmholtz coils were placed inside a magnetically shielded cylinder comprising 7 layers of μ-metals. Fig. 3(a) shows the output signal as a function of $\mu_0 H_y$ in the range of ±20 μT for the full bridge AMR sensors with $a$ = 800 μm (solid-line), 400 μm (dashed-dotted-line) and 200 μm (dotted-line), at bias current densities of $j_{Pt} = 7.34\times10^{5}$ A $cm^{-2}$, $9.18\times10^{5}$ A $cm^{-2}$, $1.65\times10^{6}$ A $cm^{-2}$, respectively. As can be seen from the figure, all the three sensors exhibit good sensitivity and linearity. Before we characterized the bridge sensor, we investigated how each sensing element responds to external field at different biasing current density. We confirmed that the response curve of individual sensing element of each sensor exhibits a field shift with its sign dependent on the bias current direction (not shown here);

this excludes Joule heating as the cause for the observed MR response.

From the slope of the response curve, one can extract the sensitivity of each sensor and the results are tabulated in Table I together with other performance indicators. The sensitivity of the sensors is comparable to that of commercial AMR sensors.[36] The sensor exhibits good linearity at low field, but the linearity error increases at high field. To best characterize this relationship, we show in Fig. 3(b) the linearity error as a function of the dynamic range. Here, the linearity error (%) is the deviation of the sensor output curve from a specified straight line over a desired dynamic range. The general trend is that the linearity error increases as the dynamic range increases, which is typical for AMR sensors. By defining the working field range as the dynamic range that gives a linearity error below 5%, we obtain the field range for the three sensors and the values are also listed in Table I. From these results, we can see that by changing the ellipsoid dimension, we can tune the sensor's working field range and sensitivity via manipulation of shape anisotropy. Compared to commercial AMR sensors, the dynamic range of SOT-biased NiFe/Pt sensor demonstrated in this work is relatively small. This is mainly because of the fact that, in this specific material combination, $H_{FL}/j_{Pt}$ is around $0.7\times10^{-6}$ Oe ($A^{-1}$ $cm^2$) and the current density used is on the order of $10^6$ A $cm^{-2}$. If we increase the current density to $10^7$ A $cm^{-2}$, we should be able to increase the dynamic range to be comparable with that of commercial sensors. The current density can be reduced further without sacrificing the dynamic range when FM/HM heterostructures with large SOT effective field are found in future.

In order to examine the detection limit of these SOT-biased AMR sensors, we performed AC field sensing experiments and analyzed the waveform of the output signal. In these experiments, an AC magnetic field with different magnitudes and fixed frequency of 0.1 Hz was applied in *y*-direction, and the sensor output was recorded with respective to time. As an example, Figs. 4(a) – 4(c) show the output signal for the sensor with $a$ = 800 μm. The sensors are biased at a current density of $j_{Pt} = 7.34 \times 10^5$ A $cm^{-2}$ and the amplitudes of the AC field are 30 μT, 500 nT and 20 nT, respectively. The amplitude of output signal decreases with the amplitude of applied field, and is eventually masked out by the noise when the latter is below 20 nT. To have a better understanding of the detection limit, Fourier transform (FT) of the output waveforms is performed, and the corresponding results are shown in Figs. 4(d) – 4(f). As can be seen, a clear peak at 0.1 Hz can be identified for all cases. However, as the applied field amplitude decreases further to below 20 nT, the peak becomes indiscernible (not shown here). Therefore, the detection limit of the sensor with $a$ = 800 μm is around 20 nT. Similar measurements were performed on the other two sensors and the detection limits are given in Table I. The increase of dynamic range is realized through size reduction which gives a larger demagnetizing field. As this will also make the sensor less responsive to the external field, it leads to a lower sensitivity. For practical applications, the dimension of the sensor can be optimized based on the application requirements. In Table I, we also include the power consumption for the sensor element, which is comparable to that of the commercial sensors.

In summary, we demonstrated a semitransparent AMR/SMR sensor enabled by SOT-biasing in NiFe/Pt heterostructures. Despite its ultrathin thickness and extremely simple design, the full Wheatstone bridge sensor prepared exhibits performances comparable to those of commercial sensors using more complicated biasing schemes. The simple structure and semitransparency will expand the range of applications of magnetic sensors in IOT and smart living. We hope this work will stimulate more follow up efforts on the development of robust, cheap and transparent AMR/SMR sensors based on novel spin-orbit physics phenomena.

## Acknowledgements

Y.H.W. would like to acknowledge support by the Singapore National Research Foundation, Prime Minister's Office, under its Competitive Research Programme (Grant No. NRF-CRP10-2012-03) and Ministry of Education, Singapore under its Tier 2 Grant (Grant No. MOE2013-T2-2-096). Y.H.W. is a member of the Singapore Spintronics Consortium (SG-SPIN).

## FIGURE CAPTIONS

FIG. 1. (a) Simulated transmittance in the visible range for NiFe(1.5)/HM(2) bilayers on quartz substrate with different HM: Pt (solid-line), Ta (dashed-line) and W (dotted-line); (b) Simulated oxide thickness dependence of transmittance at $\lambda$ = 500 nm for Pt(2)/NiFe(1.5)/oxide($d_{oxide}$) trilayers with different oxides: $Ta_2O_5$ (dashed-dotted-line), $SiO_2$ (dotted-line), MgO (solid-line), $Al_2O_3$ (short-dashed-line) and bilayer without oxide (dashed-line); (c) Measured transmittance for NiFe(1.5)/Pt($d_{Pt}$) with $d_{Pt}$ = 1.5 nm (dashed-dotted-line), 2 nm (solid-line) and 2.5 nm (dotted-line) and the bare substrate (dashed-line); (d) Measured *M-H* loops for NiFe($d_{NiFe}$)/Pt(2) films with $d_{NiFe}$ = 1.7 nm, 1.8 nm, 1.9 nm and 2 nm. Insets of (a) and (b) are the schematics of NiFe/HM bilayer and Pt/NiFe/Oxide trilayer, respectively; and inset of (c) is the photograph of NUS logo covered by the coupon film of NiFe(1.8)/HM(2).

FIG. 2. (a) Schematic of a full Wheatstone bridge AMR sensor consisting of four ellipsoidal NiFe/Pt bilayer sensing elements with the arrows indicating the magnetization direction biased by the SOT effective field; (b) Illustration of the SOT-biasing scheme; (c) Experimentally determined $H_{FL}$ values at different $j_{Pt}$ values using 2$^{nd}$ order PHE measurements; (d) Calculated $\varphi$ values at different $j_{Pt}$ values. The dashed lines in (d) indicate $\varphi = \pm 45^{o}$.

FIG. 3. (a) Typical output signal as a function of $\mu_0 H_y$ in the range of ±20 μT for the three full bridge sensors with $a$ = 800 μm (solid-line), 400 μm (dashed-dotted-line) and 200 μm (dotted-line); (b) Summary of the linearity error at different dynamic ranges for the three sensors.

FIG. 4. (a) – (c) Output signal for the sensor with $a$ = 800 μm and at $j_{Pt} = 7.34 \times 10^5$ A cm$^{-2}$, subjecting to an external AC field with different amplitude: 30 μT, 500 nT and 20 nT; (d) – (f) Fourier transform of the waveforms in (a) – (c).

TABLE I A summary of the sensor performance parameters with different dimensions. FS is the abbreviation of full scale.

| Dimension (μm) | Sensitivity (mΩ $Oe^{-1}$) | Field range (μT) | Linearity error (% FS) | Power (mW) | Detection limit (nT) |
|---|---|---|---|---|---|
| 800 × 200 | 202.9 | -30 ~ 30 | 0.9 (± 0.05 Oe)<br>2.5 (± 0.2 Oe)<br>4.8 (± 0.3 Oe) | 3.60 | 20 |
| 400 × 100 | 139.2 | -50 ~ 50 | 1.8 (± 0.05 Oe)<br>3.1 (± 0.2 Oe)<br>4.9 (± 0.5 Oe) | 1.44 | 50 |
| 200 × 50 | 169.3 | -80 ~ 80 | 1.5 (± 0.1 Oe)<br>3.2 (± 0.4 Oe)<br>5.0 (± 0.8 Oe) | 1.17 | 200 |

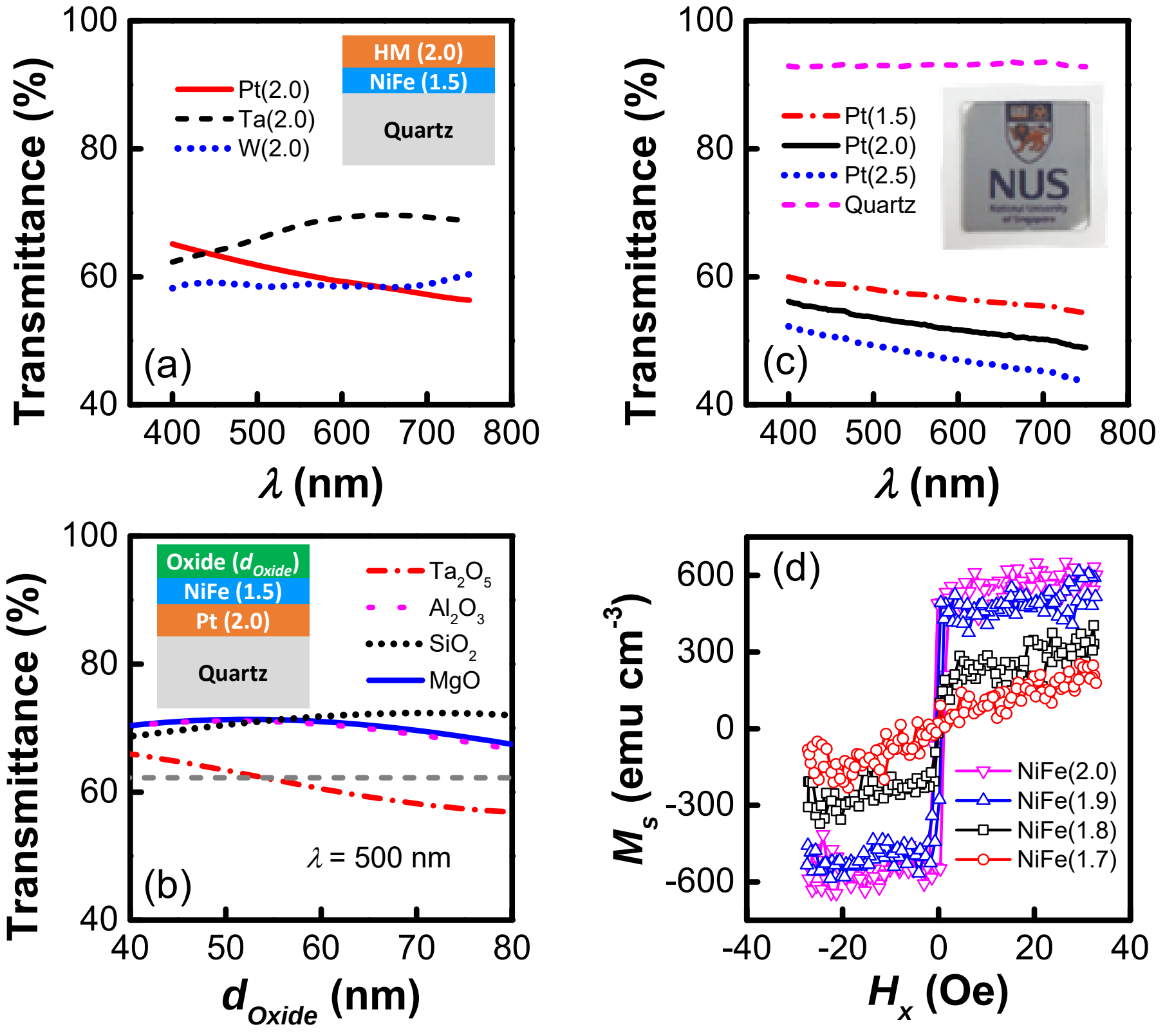

FIG. 1

Applied Physics Letters

Yumeng Yang

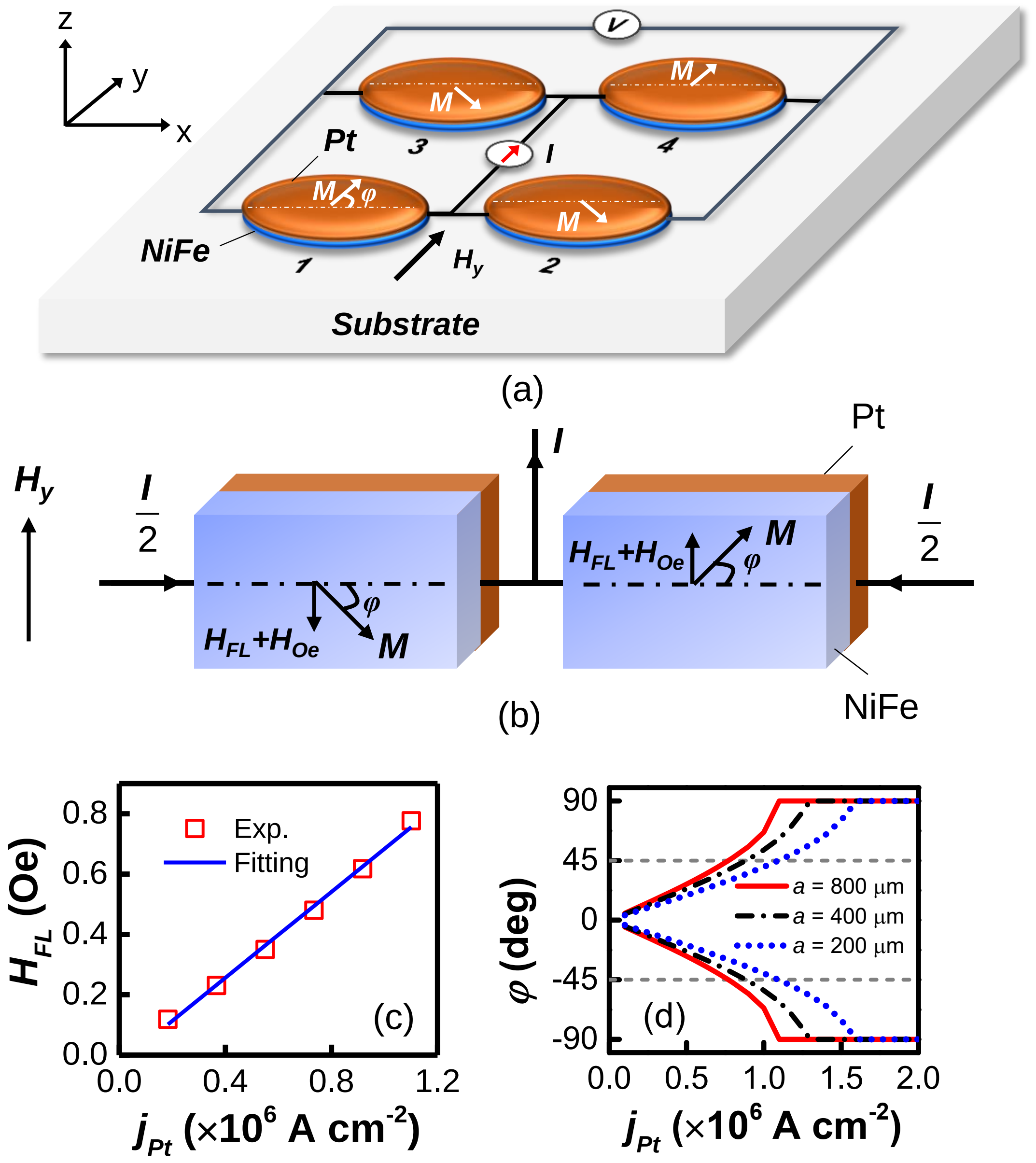

FIG. 2

Applied Physics Letters

Yumeng Yang

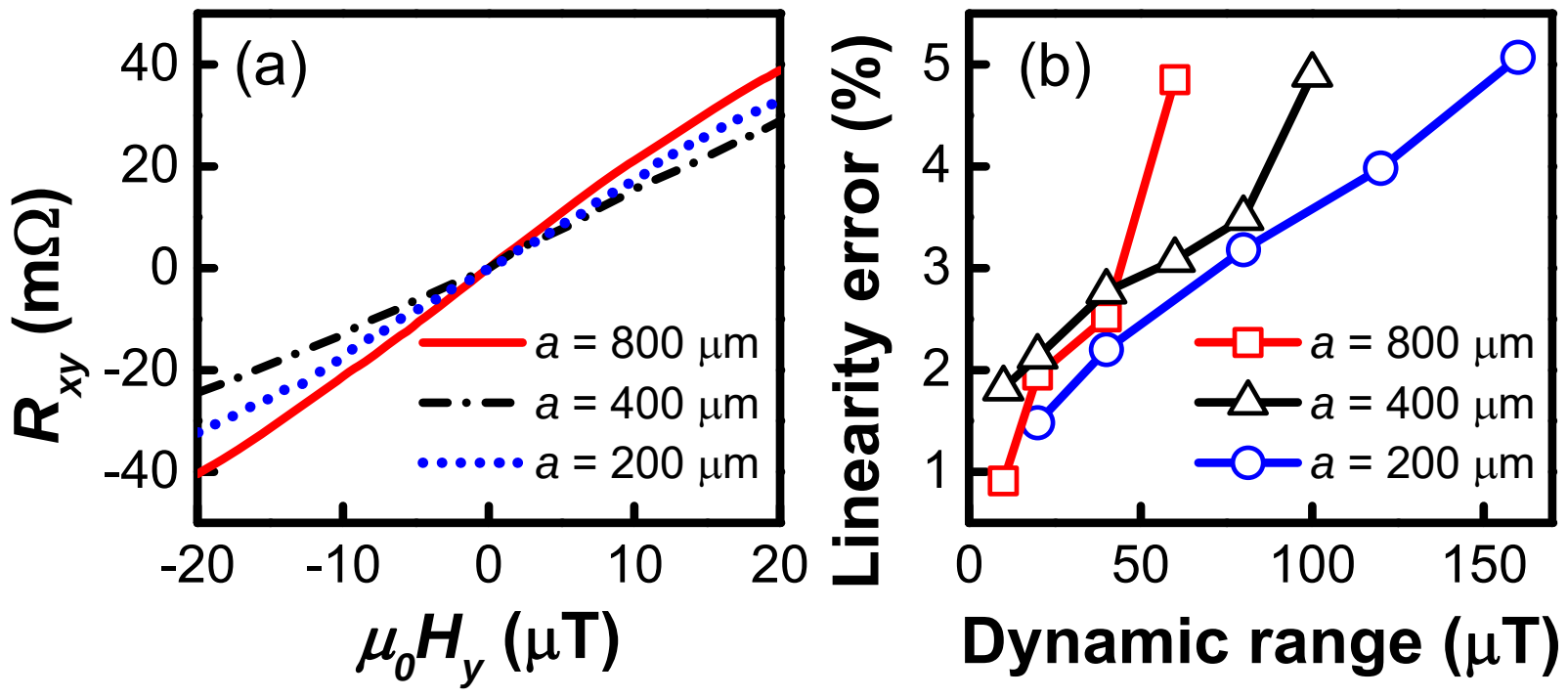

FIG. 3

Applied Physics Letters

Yumeng Yang

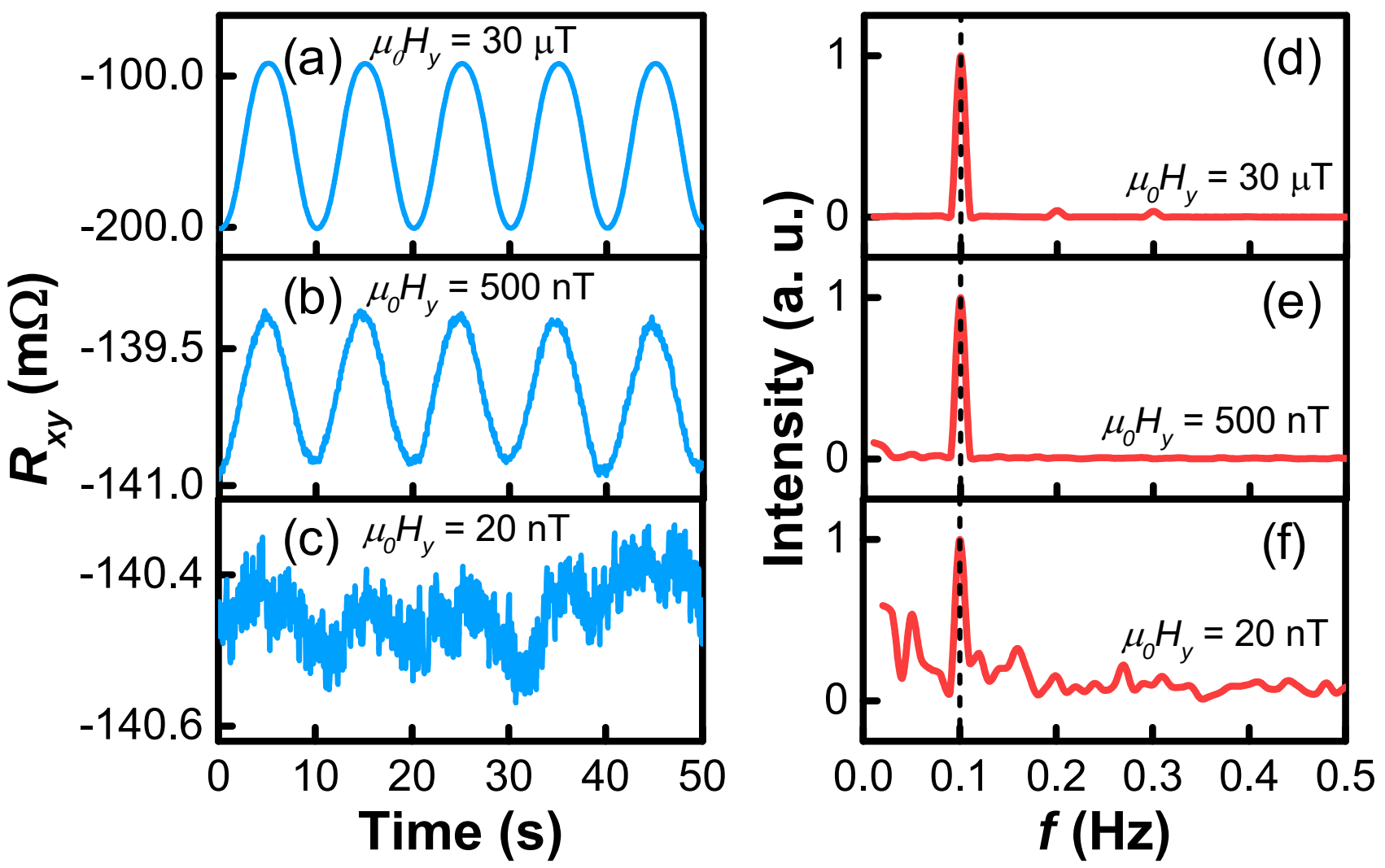

FIG. 4

Applied Physics Letters

Yumeng Yang